\documentclass [aps, pra, twocolumn, oneside, 10pt, amsfonts, amsmath, floatfix, superscriptaddress, nofootinbib] {revtex4-1}

\usepackage{amssymb}
\usepackage{etoolbox}
\usepackage{graphicx}
\usepackage{microtype}
\usepackage{physics}
\usepackage{scalerel}
\usepackage{tikz}
\usepackage{xcolor}
\usepackage{hyperref}

\pdfoutput=1
\makeatletter
\renewcommand\Dated@name{}
\makeatother

\DeclareMathOperator\erfc{erfc}

\usetikzlibrary{svg.path}

\definecolor{orcidlogocol}{HTML}{A6CE39}
\tikzset{
  orcidlogo/.pic={
    \fill[orcidlogocol] svg{M256,128c0,70.7-57.3,128-128,128C57.3,256,0,198.7,0,128C0,57.3,57.3,0,128,0C198.7,0,256,57.3,256,128z};
    \fill[white] svg{M86.3,186.2H70.9V79.1h15.4v48.4V186.2z}
                 svg{M108.9,79.1h41.6c39.6,0,57,28.3,57,53.6c0,27.5-21.5,53.6-56.8,53.6h-41.8V79.1z M124.3,172.4h24.5c34.9,0,42.9-26.5,42.9-39.7c0-21.5-13.7-39.7-43.7-39.7h-23.7V172.4z}
                 svg{M88.7,56.8c0,5.5-4.5,10.1-10.1,10.1c-5.6,0-10.1-4.6-10.1-10.1c0-5.6,4.5-10.1,10.1-10.1C84.2,46.7,88.7,51.3,88.7,56.8z};
  }
}

\newcommand\orcidicon[1]{\href{https://orcid.org/#1}{\mbox{\scalerel*{
\begin{tikzpicture}[yscale=-1,transform shape]
\pic{orcidlogo};
\end{tikzpicture}
}{|}}}}

\begin {document}

\title {Anderson Localization on the Bethe Lattice using Cages and the Wegner Flow}

\author {Samuel \surname {Savitz} \orcidicon{0000-0003-2112-3758}}\email {Sam@Savitz.org}
\affiliation {Institute for Quantum Information and Matter, Department of Physics, California Institute of Technology, Pasadena,~California~91125,~USA}
\author {Changnan \surname {Peng} \orcidicon{0000-0002-9331-2614}}\email {CnPeng@MIT.edu}
\affiliation {Institute for Quantum Information and Matter, Department of Physics, California Institute of Technology, Pasadena,~California~91125,~USA}
\affiliation {Department of Physics, Massachusetts Institute of Technology, Cambridge, Massachusetts 02139, USA}
\author {Gil \surname {Refael}}\email {Refael@Caltech.edu}
\affiliation {Institute for Quantum Information and Matter, Department of Physics, California Institute of Technology, Pasadena,~California~91125,~USA}

\date {Uploaded April 15\textsuperscript{th}, 2009; Published September 16\textsuperscript {th}, 2019}

\begin {abstract}
    Anderson localization on tree-like graphs such as the Bethe lattice, Cayley tree, or random regular graphs has attracted attention due to its apparent mathematical tractability, hypothesized connections to many-body localization, and the possibility of \emph {non-ergodic extended} regimes.  This behavior has been conjectured to also appear in many-body localization as a ``bad metal'' phase, and constitutes an intermediate possibility between the extremes of ergodic quantum chaos and integrable localization.  Despite decades of research, a complete consensus understanding of this model remains elusive.  Here, we use cages, maximally tree-like structures from extremal graph theory; and numerical continuous unitary Wegner flows of the Anderson Hamiltonian to develop an intuitive picture which, after extrapolating to the infinite Bethe lattice, appears to capture ergodic, non-ergodic extended, and fully localized behavior.
\\\\
\href {https://doi.org/10.1103/PhysRevB.100.094201}{Physical Review B \textbf{100}, 094201 (2019).  DOI:  10.1103/PhysRevB.100.094201}
\\[8pt]
Physics Subject Headings (PhySH):  \emph{Condensed Matter \& Materials Physics}
\begin{description}
\item[Research Areas] {Electronic structure, Localization, \underline{Metal--insulator transition}, Quantum chaos,\\ Quantum phase transitions}
\item[Physical Systems] {High dimensional systems, Quantum chaotic systems}
\item[Techniques] {Random matrix theory, Tight-binding model}
\end {description}
\end {abstract}

\maketitle

\section {Introduction} \label {Introduction}

Localization due to quenched disorder has been a major paradigm in quantum mechanics since its introduction by Anderson in 1958~\cite{Anderson58}.  When a wave travels between two distinct media, some fraction is usually reflected.  When the media is disordered, these reflections can destructively interfere, leading to the observation of Anderson localization.

Disorder is ubiquitous in nature, and the non-ergodic behavior of localized systems has potentially crucial or even practically useful implications for the process of thermalization.  For example, non-thermalizing phenomena could be used to reduce decoherence in a quantum computer~\cite{Nandkishore15,Altman15}, or cause an adiabatic quantum computer to be very slow~\cite{Altshuler10,Suzuki15}.  Despite the intensive study of localization, there are still many cases where ``one has to resort to the indignity of numerical simulations to settle even the simplest questions about it.''~\cite{Anderson78}  The most important known exceptions to this generalization are settings in which there are no spatial ``loops''.  For example, single-particle Anderson localization on an infinite one-dimensional chain can be treated analytically using transfer matrices.

Another more complicated loopless setting is a single particle on the infinite Bethe lattice tree.~\cite{Bethe35}   This problem was first ``solved'' by Abou-Chacra, Anderson, and Thouless (ACAT) in 1973.~\cite{Abou-Chacra73,Abou-Chacra74} They found that for a disorder strength above $W_c \approx 18$, all eigenstates are localized.  Recently this model, the focus of this work, has attracted significant attention due to its proposed connections to \emph {many-body} localization (MBL) which is discussed in section~\ref {MBL}.  More recent work indicates that the ACAT solution does not tell the whole story.  Many diverse and advanced techniques have been applied to Anderson localization on the Bethe lattice, suggesting sometimes contradictory phase diagrams.  We seek to further the understanding of this model using the relatively elementary numerical technique of Wegner--Wilson flows, which provides a particularly intuitive picture of the localization phenomenon.

\begin{figure} [t]
    \begin {center} 
	\includegraphics{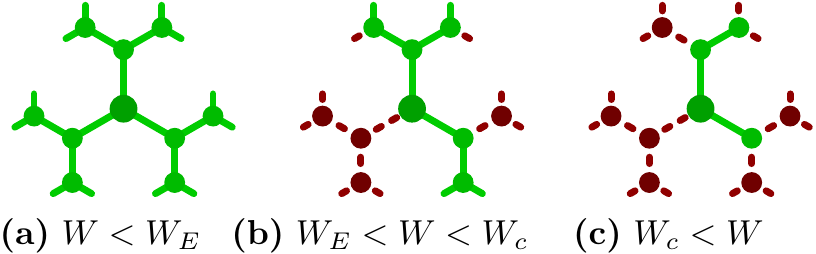}
    \end{center}
    \vspace*{-10pt}
    \caption{
        \textbf {Schematic Depiction of the Three Hypothesized Phases:}\\
        \protect\hphantom{---}The focus of this work is quantum mechanical particles hopping on these tree graphs (the Bethe lattice) with on-site Anderson disorder.\\
        \protect\hphantom{---}In \textbf {(a)}, ergodic states for sufficiently small disorder are shown as extended over the entire tree.  On the Bethe lattice, the resulting dynamical diffusion is ballistic.  At the edges of the spectrum, non-ergodic states are expected to appear, but there may be a level of disorder below which localization never occurs, due to resonant delocalization~\cite{Aizenman11,Warzel12,Aizenman13}.\\
        \protect\hphantom{---}In \textbf {(b)}, a non-ergodic extended phase is shown.  Note that the wavefunction density is highest at the central node and decays exponentially as one moves away from it regardless of whether the path taken is green or dark red.  However, the exponentially rare, but also exponentially numerous green paths decay slowly enough that the state can not be normalized.  This is discussed in more mathematical detail in section~\ref{Picture}.\\
        \protect\hphantom{---}Finally, in \textbf {(c)}, the disorder is sufficiently strong that all states are exponentially localized around a central site.
    } \hrule \vspace*{-5pt} \label{Trees}
\end{figure}

The trivalent Bethe lattice, shown in figure~\ref{Trees}, can be characterized as the only infinite, connected, cycle-free, cubic graph.  It is the homogeneous infinite tree graph with coordination number three.  The Anderson localization Hamiltonian can be written as
\begin {equation}
    H = \sum_a \epsilon_a \dyad{a} - \sum_{\langle ab \rangle} \dyad{a}{b}, \label{Anderson}
\end {equation}
where the $\epsilon$'s are drawn from the uniform distribution with support $[-W/2, W/2]$, and the second sum is taken over ordered pairs of connected nodes on our graph.  One is generally interested in the spectrum of this operator as a function of $W$ and the structure of its eigenstates.  The structure of the eigenstates depends on their eigenvalue.  For example, a so-called energetic ``mobility edge'' can separate localized from extended states.

For the basic Anderson model given in equation~\eqref {Anderson}, on one- and two-dimensional lattices, any amount of quenched disorder is sufficient to localize a particle.  On higher-dimensional lattices, weak disorder admits extended eigenstates, but for $W$ above some critical $W_c$, all eigenstates are localized.  When $W$ is exactly tuned to $W_c$, the eigenstates adopt an intricate \emph {multifractal} structure.~\cite{Abrahams79,Jia08,Elihu10,Evers08,Lindinger17}

As mentioned above, the literature on the problem of localization on the Bethe lattice has not yet reached a clear consensus, and our contribution is unlikely to be the final word on the matter.  Our work finds evidence for a particularly common understanding of the phase diagram and we will present intuitive arguments in its favor, but the reader should keep in mind that the issue remains controversial, and our results are by no means conclusive.  An incomplete summary of alternative claims will be presented in section~\ref {Review}.

With that caveat, on the Bethe lattice, the $W = W_c$ point of multifractal behavior expands into a region $W \in \left[ W_E, W_c \right] $.  For $W < W_E$, there is an absolutely continuous spectrum of non-normalizable states extended over the entire tree.  For $W > W_c$, all states are localized around a central node and normalizable, per the results of ACAT.  Finally, for the intermediate regime $W_E < W < W_c$, the eigenstates are thought to be extended but non-ergodic.  This means that they occupy an infinite number of sites in a non-normalizable fashion, but extend over an only infinitesimal fraction of the tree.  The spectrum of these states is thought to be still absolutely continuous, but we believe the eigenstates as a function of their energy are discontinuous over arbitrarily small changes in eigenvalue in the sense of having very different branching patterns.  By ``branching pattern'', we are referring to the subset of occupied sites such as shown in figure~\ref{Trees}(b).  These critical disorder values, $W_E$ and $W_c$, are defined for states in the band center, \emph{i.e.}~$E \approx 0$, where they are the largest.  This neither localized nor ergodic intermediate behavior is particularly interesting because MBL has also been suggested to exhibit non-ergodic extended behavior in Fock space over a finite region of its phase diagram~\cite{DeLuca13,BarLev15,Agarwal15,Monthus16b,Luitz16,Znidaric16,Torres-Herrera17,Luitz17}.

Intuitively, we can think of these three cases as follows:  In the ergodic extended regime, particles which are reflected by the disorder have many options regarding where to propagate next.  In one dimension they can only return to their previous location, but on the tree, their random walk quickly takes them ``very far from home''.  One should think of the tree as more akin to an infinite-dimensional lattice than to a one-dimensional structure.  In the localized phase, however, the reflection is so strong that the particle still never manages to wander far from its starting location.  For non-ergodic extended states, we can think of the success of the particle traveling down a branch as a probabilistic process.  At each new node it encounters two new branches, and while it doesn't always propagate through both, it has an expected value of extending through more than one.  This leads to the state occupying infinitely many nodes, a number which grows exponentially with distance, but an exponentially small fraction of the graph.  A caricature of these three phases is shown in figure~\ref{Trees}.
While a useful approximation, thinking of a particle's propagation or reflection as a binary result does not fully capture the structure of the eigenstates.  A more rigorous explanation of this intuition is presented in section~\ref{Picture}.

Numerically, we approached this matrix diagonalization problem using the technique of Wegner flows, the most common type of continuous unitary transformation.  While flows are not the most efficient way to diagonalize a matrix, they provide a pleasing temporal picture of the construction of the diagonalizing unitary:  Hopping bonds joining sites at differing energies appear to decay while causing those energy levels to further repel and generating new, longer-range bonds.  Furthermore, one can keep track of this level repulsion using the $\Xi$ metric.  For more details, see section~\ref{Flows}.~\cite{Wegner94,Glazek93,Glazek94,Savitz17}

Of course, the infinite Bethe lattice can not be simulated on a finite computer.  As discussed in section~\ref{Review}, many finite analogs of the Bethe lattice have been used in previous research.  In this work, we will adopt the graph theory concept of ``cages'' to provide a finite setting which is in some sense optimally similar to the Bethe lattice.  Briefly, cages are degree-regular graphs with a minimal number of sites given their ``girth''.  A graph's girth $g$ is the length of its shortest loop.  Thus, a regular graph always ``looks like'' the Bethe lattice until one has taken $g$ hops away from any initial site.~\cite{Tutte47,Erdos63}  We are using the smallest known cubic graph of girth eighteen which was discovered by Exoo and has 2560 sites~\cite{Exoo08}. For more details on cages, see section~\ref{Cages}.

Finally, we conclude with our numerical results in section~\ref{NumericalResults}, and a discussion of further directions and connections in section~\ref{Discussion}.

\section {Connection to Many-Body Localization} \label{MBL}
As mentioned above, any amount of quenched disorder is sufficient to localize a single particle in one dimension.  It was long believed that interactions between multiple particles would cause them to become delocalized.  However, Basko, Aleiner, and Altshuler showed that the behavior of interacting disordered particles is more complicated.~\cite{Basko06,Altshuler97}  For sufficiently weak disorder, the interactions do indeed delocalize the particles into an ergodic fluid, but for sufficiently strong disorder, all of the eigenstates of band-limited systems can in fact be localized.  Characterizing this ``many-body localization'' transition has been a major goal of condensed matter physics in the past decade.~\cite{Nandkishore15,Altman15} 

It is hoped that progress in this effort would shed light on the fundamental difference between ergodic quantum chaos and integrable localized behavior.~\cite{Srednicki94,Serbyn13,Huse14,Rademaker17} The localized phase may be useful as a persistent quantum memory~\cite{Nandkishore15}, and its non-thermalizing behavior may have important thermodynamic implications for nanodevices such as quantum engines~\cite{Halpern19}. 

One major difficulty with this goal lies in the lack of experimental and numerical data with which to compare theoretical results. Experimental realizations of localized systems usually require extreme isolation from thermal baths, and, at least na\"{\i}vely, numerical simulations require exponentially large amounts of memory and computational expense as a function of the system size.  Many suspect that the 20--30-site numerical results currently available suffer from finite-size effects which prevent the true scaling limit of the MBL transition from being observed.~\cite{Dumitrescu17} 

A second difficulty lies in the apparent ``fuzziness'' of the transition between the ergodic and localized behavior.  Some have assumed that this is a result of the finite-size effects~\cite{Oganesyan07,Chandran15}, but others have argued for a more complicated phase diagram with at least one intermediate phase, which is neither ergodic nor entirely localized~\cite{DeLuca13,Grover14,Gopalakrishnan15,BarLev15,Agarwal15,Serbyn16,Luitz16,Torres-Herrera17,Luitz17}.  This non-ergodic extended phase has been referred to as that of a ``bad metal''~\cite{Pino16}, which allows for some amount of current to flow, but only through a vanishingly small fraction of Fock space.  This three-phase behavior is shared with our understanding of the disordered Bethe lattice model, where the multifractal nature of the non-ergodic extended states have a particularly straightforward interpretation.~\cite{DeLuca14,Sonner17,Kravtsov18}

One can derive the disordered Bethe lattice model from that of MBL through three non-rigorous logical steps.  First, MBL can be represented without any approximation as a single particle hopping on a disordered $N$-dimensional Fock space, where $N \to \infty$ is the number of particles.~\cite{Gornyi05,Monthus10}  The on-site disorder of the high-dimensional hypercubic lattice is spatially correlated.  The first, substantial, and not rigorously justified, approximation is to assume that this correlation is irrelevant after sufficient renormalization group flow towards the scaling limit.~\cite{DeLuca13}  The upper critical dimension of Anderson localization has been debated in the literature, but some suggest it to be this infinite-dimensional limit.~\cite{Mirlin94,Tarquini17}  On an infinite-dimensional hypercubic lattice, random walkers almost never return to their initial position, so loops are negligible, and in the second approximation, we can model the infinite-dimensional hypercubic lattice as a Bethe lattice with infinite coordination number.  One notable difference between these settings is the distance metric.  Because the pythagorean theorem is no longer applicable as on a Euclidean lattice, diffusion on the tree is inherently ballistic.  The final approximation is to reduce the coordination number to the minimal non-linear option of three.  The qualitative results appear to be independent of the finite coordination number so long as the graph is tree-like, so one can extrapolate from the cubic case to higher-coordination trees.~\cite{Kravtsov18}  This, along with replacing the tree with a cage, is done to make the problem numerically tractable, hopefully without sacrificing the most salient features of MBL.

The Bethe lattice model has the clear advantage of being loopless, allowing for many analytic approaches to succeed with fewer approximations than the usual toy models of MBL, which have to contend with complicated many-particle configurations.  The exponentially large Hilbert spaces of many-body states are exchanged for a single particle in an exponentially vast spatial setting.  Additionally, unlike many-body configurations, two sites on the tree have a clear and relevant definition of distance, and we will utilize this structure in the analysis in section~\ref{Picture}.

\section {Review of Previous Results} \label{Review}
Many analytic and numerical techniques have been applied to the model of Anderson localization on the Bethe lattice.  It and other high-dimensional Anderson localization models have been reviewed in Tarquini's thesis~\cite{Tarquini16b}.  While we can not hope to be comprehensive, we will attempt to provide a succinct summary of the relevant literature.

Analytic approaches include the original self-consistent Green's function distributions of Abou-Chacra \emph{et al.}~(although their approach requires numerics to find a value for $W_c$)~\cite{Abou-Chacra73,Abou-Chacra74}, dynamical evolution~\cite{Klein96,Aizenman12}, multifractal analysis~\cite{DeLuca14}, supersymmetric non-linear $\sigma$-models~\cite{Zirnbauer86,Verbaarschot88,Mirlin91,Tikhonov19} (see~\cite{Efetov99} for an introduction to the application of supersymmetry to disordered systems), and replica symmetry breaking~\cite{Klein98,Kravtsov18}.

Many numerical methods have also been applied to this model.  One common option is the spectral level spacing statistic, or $r$-parameter~\cite{Biroli12,Tikhonov16b}, and the level compressibility has also been probed~\cite{Metz17}.  The (multi)fractal nature of the non-ergodic eigenstates has quantified using the inverse participation ratio and its generalizations.~\cite{Biroli12,Sonner17}  Other numerical diagnostics include the return probability~\cite{Bera18}, transmission through branching wires~\cite{Monthus11}, imbalance, and the two-point equilibrium dynamical correlation function~\cite{Biroli17}.  Dynamical propagation has also been performed numerically, finding subdifussive behavior for certain disorder strengths~\cite{Tomasi19}, in contrast to the ballistic behavior predicted analytically for weak disorder~\cite{Klein96} or when the spectrum is absolutely continuous~\cite{Aizenman12}.

In the non-spatial Rosenzweig--Porter random matrix model~\cite{Rosenzweig60}, which has also been suggested to be connected to many-body localization~\cite{Shukla16}, the non-ergodic extended states~\cite{Kravtsov15,Facoetti16,Monthus17,vonSoosten19,Pino19} have Wigner--Dyson level spacings and Gaussian ensemble $r$-parameter values in the large-$N$ scaling limit.  This is because, while the Thouless energy does decay to zero with increasing $N$, the mean level density increases at an asymptotically higher rate.~\cite{Kravtsov15}\hyperlink{fn}{*}\footnotetext{It is interesting to note that, independent of the value of $W > 0$, scrambling the off-diagonal elements of the Bethe lattice's Anderson localization Hamiltonian while keeping their Frobenius norm constant gives the $\gamma = 1$ ergodic--non-ergodic extended transition point of Rosenzweig--Porter random matrix model.}  It is important to note that, contrary to the Rosenzweig--Porter random matrix case, two non-ergodic extended states on the Bethe lattice are expected to have negligible overlap, and should therefore exhibit Poissonian level statistics and $r$-parameter.~\cite{Biroli12}  Another non-spatial model with apparent non-ergodic extended states at small length scales are random L\'evy matrices.~\cite{Tarquini16a}

Spatial systems thought to also exhibit non-ergodic extended behavior include disordered Josephson junction chains~\cite{Pino16,Pino17} and the Aubry--Andr\'e model~\cite{Li15}.

Numerical approaches must choose a finite version of the Bethe lattice, and this can be a source of subtle discrepancies.~\cite{Ostilli12}  Simple truncation~\cite{Sade03,Sonner17} suffers from having a large fraction of the sites near the leaves of the tree.~\cite{Tikhonov16a,Biroli18}  Random regular graphs have been proposed as a good alternative~\cite{Biroli12,DeLuca14,Altshuler16b,Tikhonov16b,Biroli17,Metz17,GarciaMata17,Kravtsov18,Tikhonov18,Bera18,Tikhonov19,GarciaMata19,Tikhonov19b}, as they are expected to look locally tree-like.  However, for numerically realistic sizes, small loops are inevitable and represent a source of additional, topological disorder.~\cite{Kravtsov18}  Furthermore, the expected global girth actually converges to a finite value and even triangles occur with finite probability.~\cite{Wormald78,McKay04}  While these small loops are almost certainly irrelevant in the scaling limit, they are not necessary and can cause subtle difficulties when trying to define the distance between two sites of a random regular graph.  Cages seem to represent an optimal solution to this problem, as sites at a distance $d \ll g/2$ always have a clear integer distance given by the length of the only short path that connects them.  Finally, the interpretation of data from finite systems is complicated by the possibility of crossover effects at very large sizes.~\cite{Tikhonov16a,Tikhonov16b,Biroli17,Biroli18,Metz17,Tikhonov19,Tikhonov19b}

Other related geometries that have been studied include \hypertarget{fn}{a} truncated Bethe lattice with random links connecting the leaves~\cite{Sade03}, an incoming wire branching into a binary tree, \emph{i.e.}~the Miller--Derrida scattering geometry~\cite{Monthus11}, a one-dimensional chain with random shortcuts~\cite{GarciaMata17}, and metric trees~\cite{Damanik19,Aizenman06a}.

Using early numerics, ACAT estimated the critical disorder strength above which all states are localized, $W_c$, to be about 18.~\cite{Abou-Chacra73,Abou-Chacra74}.  Subsequent values that have been suggested, using methods with variable degrees of approximation, include 16.99--17.32~\cite{Monthus09}, $17 \pm 1$~\cite{Monthus11}, 17.4~\cite{Biroli10}, 18.00--18.61~\cite{Bapst14}, $18.1 \pm 0.5$~\cite{Altshuler16b}, 17.65 or $18.6 \pm 0.3$~\cite{Kravtsov18}, $18.11 \pm 0.02$~\cite{Parisi18}, and $18.17 \pm 0.01$~\cite{Tikhonov19b}.  We find $W_c$ values which are slightly below these, but do not expect our approach to be quantitatively precise.

The critical disorder strength above which all states are non-ergodic, $W_E$, is more difficult to estimate.  It has been suggested to be approximately 10~\cite{Altshuler16b}, 5.74~\cite {Kravtsov18}, and bounded between $0.4 \, W_c$ and $0.7 \, W_c$~\cite{Bera18}.  By isolating the data coming from the center of the spectral band, we find a small but positive $W_E \approx 2.1$ which we interpret as confirming the existence of a fully ergodic phase with sufficiently weak disorder.

Given the wide variety of analytical and numerical techniques and graph settings, it can be difficult to ascertain the literature consensus.  In some cases the results appear to be contradictory, and it is not immediately obvious whether this is due to finite-size effects, details of the setting, or a mathematically difficult but invalid approach.  Many potential phase diagrams have been proposed, including options which deny the existence of ergodic states at positive disorder.  Additionally, the intermediate non-ergodic extended phase was not originally anticipated by ACAT,~\cite{Abou-Chacra73,Abou-Chacra74} and this has led to some confusion.

Many authors believe that the intermediate non-ergodic extended phase eventually crosses over to ergodic behavior at large length scales.~\cite{Tikhonov16a,Tikhonov16b,Biroli17,Biroli18,Metz17,Tikhonov19,Tikhonov19b}  This may be a consequence of the use of the graphs with loops as a finite analogue of the Bethe lattice, as multifractality on the truncated Bethe tree is less controversial.~\cite{Monthus11,Tikhonov16a,Biroli18}  We hope that the extrapolation applied below, enabled by our use of cages with a clear distance metric, makes our results more applicable to the truly infinite and loopless Bethe lattice, and this may explain the apparent discrepancy between these results and our own.  Conversely, it has also been claimed that the entire extended region is non-ergodic~\cite{DeLuca14}, in contradiction to analytic results~\cite{Anantharaman17b,Anantharaman17a}.  Finally, a significant contingent agrees with the three-regime phase diagram presented in this work, at least under the correct conditions.~\cite{Biroli12,Altshuler16b,Kravtsov18,Bera18}

We hope our relatively elementary flow approach and resulting intuitive picture of the three-phase proposal has a clarifying effect on this confusing situation.

\section {Wegner Flows} \label{Flows}

Wegner flows are the best-known continuous unitary transformation process.  The Wegner flow parameterizes the diagonalization process along an artificial time dimension $\tau$.  At $H(\tau = 0)$, the Hamiltonian is expressed in its original position-space basis.   As the flow progresses, a series of infinitesimal unitary changes of bases are applied to the Hamiltonian matrix, each of which brings it closer to a diagonalized form.  For almost every initial Hamiltonian, the flow proceeds until the Hamiltonian $H(\tau \to \infty)$ has reached a diagonal fixed point.
Continuous unitary flows have been used to study the phenomenon of many-body localization.~\cite{Thomson18,Kelly19,Savitz17}  To learn more about Wegner flows and other continuous unitary transforms, see~\cite{Wegner94,Glazek93,Glazek94,Monthus16a,Savitz17}.

The ``equation of motion'' for the Wegner flow can be expressed as
\begin{equation}
    \pdv{H}{\tau} = \commutator{\eta}{H},
\end{equation}    
where the infinitesimal unitary rotation generator $\eta$ is $\commutator{H_\mathrm{Diag.}}{H}$, and $H_\mathrm{Diag.}$ is the diagonal part of the matrix $H$.

Continuous unitary transformations are not the most computationally efficient way to diagonalize a matrix.  However, they do provide an intuitively appealing temporal picture for the diagonalization process.  As off-diagonal elements in the Hamiltonian decay to zero, they cause the diagonal energy levels they connect to repel, and longer range hops appear on the Bethe lattice.  These too proceed to decay, iteratively generating increasingly long-range, but weaker links.  Localization on the Bethe lattice can be thought of as a competition between the generation of these longer-range hoppings, their decay, and the exponentially growing number of sites at a given distance from each node.

Although flows have been used to find local integrals of motion for MBL problems \emph{e.g.}~\cite{Rademaker17,Thomson18,Kelly19}, that is not our application.  We use flows to (almost) diagonalize a potentially dense matrix while keeping track of how much level repulsion has occured between the eigenstates associated with each pair of graph sites.
As discussed in~\cite{Savitz17}, one can monitor the level repulsion between two eigenvalues over the course of the flow using the $\Xi$ metric defined elementwise, for a real Hamiltonian, as
\begin{equation}
    \Xi_{ab} = 2 \int_0^{\tau_\mathrm {max}} \! \eta_{ab}^2 \, \mathrm d \tau,
\end{equation}
where $a$ and $b$ are matrix indices.  This can be thought of as a measure of the interaction between the two eigenstates and is closely related to the square of the Thouless energy.  Kehrein alerted us to the formal equivalence of this $\Xi$ metric and the $\infty$-Reny\'i entanglement entropy in a certain perturbative limit.~\cite{Kehrein17}  Conveniently, flows allow us to use the site to index correspondence in the original Hamiltonian matrix to then map eigenstates onto sites in a natural way, which is needed to define the distance between two eigenstates.  If one were to replace the flow with exact diagonalization and the level-repulsion metric $\Xi$ with a more traditional measure such as eigenstate overlap, they might be able to extend the calculation in section~\ref {Picture} to a higher-girth cage, but would also have to specify a natural bijection between eigenstates and sites.

We utilized a high-efficiency stabilized third-order integrator to numerically implement the Wegner flow as detailed in~\cite{Savitz17}.  By ``third-order'', we mean the accumulated error should scale approximately like the inverse cube of the number of unitary steps taken for a given $\tau_\mathrm{max}$ and error tolerance.  The stabilization allows the step size to increase at the end of the flow when the Hamiltonian is mostly diagonalized.  Each step is an exact unitary change of basis, up to floating-point rounding error.  While many previous implementations of flows, such as~\cite {Monthus16a,Thomson18,Kelly19}, drop terms deemed insignificant in order to improve efficiency, we do not.  Such truncation can perform well for localized states but tends to fail in the delocalized regime.  Our flows lead to exact diagonalization including delocalized behavior, with the only caveats being the cutoff time $\tau_\mathrm {max}$, a tolerance paramater $\epsilon$ which adaptively controls the step sizes to keep the predicted error per rotation magnitude reasonable, and double-precision floating-point rounding.  The flows were performed out to a hypothetical time of $\tau_\mathrm {max} = 2000$, which, according to the heuristics in~\cite {Savitz17}, should in general be sufficient to resolve at least $99\%$ of a coupling with a lever spacing of at least $0.024$.  At this point, the Hamiltonian is almost entirely diagonalized.  We used a tolerance parameter of $\epsilon = 5 \cdot 10^{-4}$ and, in total, consumed ${\sim} 25\,000$~CPU-hours.  Neither extending the flow nor tightening the tolerance produced qualitatively different results during tests with smaller cages.

\section {Cages} \label{Cages}
In order to avoid the pitfalls of truncating the Bethe lattice and the additional disorder and rare small loops included in random regular graphs, as mentioned in section~\ref{Review}, we utilize the concept of cages from extremal graph theory as the setting for our hopping particle.
We expect that cages have the same scaling limit as random regular graphs, but they possess a number of convenient advantages.  Formally, cages are the smallest degree-regular (trivalent or ``cubic'', in our case) graph or graphs with a given girth.  The \emph {girth} $g$ of a graph is the length of its shortest cycle.  Thus, cages are minimal in size while everywhere locally looking like the Bethe lattice.  In other words, a cage is the optimal way to connect the leaves of a truncated Bethe lattice without introducing any short loops.~\cite{Tutte47,Erdos63}

In fact, cubic $g = 18$ cages are not currently known, so we use the smallest currently available candidate discovered by Exoo~\cite{Exoo08}, which has 2560 sites.  While some cages have special symmetry properties such as vertex-transitivity, arc-transitivity, or being a Cayley graph, this one does not.  This does not pose any problem for us, because the asymmetry is only apparent when examining the graph globally.  Locally it still looks exactly like the Bethe lattice.

When studying disordered systems, one often utilizes ``disorder averaging'' to minimize the stochastic effect of a single disorder realization.  Likewise, one can use finite-size scaling, which employs extrapolation to minimize finite-size effects.  However, due to the large, self-similar structure of the Bethe lattice, we found that our computational efforts were best spent maximizing the girth of our graph.  Our cage already contains a large number of smaller trees as subgraphs, and our analysis in section~\ref{Picture} includes an extrapolation to the infinite Bethe lattice.  Additional disorder realizations have a negligible effect on our conclusions below.

The correlations of nearby sites are subject to ultraviolet details, and should be expected to be non-universal.  For example, they might depend significantly upon the artificial uniform shape of the disorder distribution.  Furthermore, sites at a distance $d \ge g/2 = 9$ can be connected by multiple minimal paths, which mutually form loops.  Thus, when analyzing our $\Xi$ level repulsion data in section~\ref{NumericalResults}, we restricted our attention to pairs of sites at distances $3 \le d \le 7$.

Performing the mapping from MBL to single-particle Anderson localization in section~\ref{MBL} without any approximations leads to a high-dimensional hypercubic graph with a girth of four as the setting.  It is therefore reasonable to ask whether our goal of maximizing the graph's girth is even desirable.  Our response is that if one wants certain answers for MBL, they should already be very cautious of the many approximations made in section~\ref{MBL}.  Regardless of how accurate they turn out to be, we take the problem of Anderson localization on the Bethe lattice as a starting point which is interesting in its own right.  Then, when selecting a finite analog for the Bethe lattice, we note that cages, while probably ultimately in the same universality class as the commonly-used random regular graphs, possess certain convenient features such as a lack of local topological disorder, thus allowing for a well-defined distance metric for separations well below the girth, which we utilize when extrapolating our results to the infinite Bethe lattice in section~\ref{Picture}.  We do not expect the MBL $g = 4$ argument above to imply that random regular graphs are \emph{superior} to cages in modeling MBL because the random regular graph's \emph{typical} loop lengths also diverge logarithmically with graph size~\cite{Tikhonov16a}.

\section {Intuitive Picture and Mathematical Analysis}\label{Picture}
\subsection{Intuitive Picture}
Schematically, our intuitive picture of the three phases is represented in figure~\ref{Trees}.  However, one should keep in mind that this ``binary'' representation of an eigenstate's support is an approximation at best.  In fact, an individual eigenstate is almost surely supported on the entirety of the tree.  In a finite number of dimensions, exponential decay of the eigenstate with distance ensures that the state is localized and normalizable.  However, on the Bethe lattice, due to the exponentially growing number of sites with distance, this is not always the case.  The non-ergodic extended eigenstates decay exponentially, but can not be normalized.  This implies that while still ``centered'' about some part of the tree, they are non-trivially supported on an infinite number of sites.  In the binary approximation, we identify some subset of the branches away from this center which are essential to the support of the eigenstate.  Similar intuitive pictures have been developed by Garc\'{\i}a-Mata \emph{et al.}~\cite{GarciaMata17,GarciaMata19}

While numerical evidence has been presented to the contrary~\cite{DeLuca14,Kravtsov15,Kravtsov18}, mathematical results suggest that for sufficiently small but positive disorder $W < W_E$, there should be an ergodic extended region of the spectrum~\cite{Anantharaman17b,Anantharaman17a,Klein94,Froese07,Klein11,Warzel12}.  This is not surprising when one remembers that the tree is effectively infinite-dimensional.  For $W_E < W < W_c$, the localization is branch-dependent.  As one moves away from the origin, some branches are sufficiently disordered to effectively localize the particle, but probabilistically, a growing number of the branches are essential components of the eigenstate's support and prevent the wavefunction from being normalized.  Again, it is important to reiterate that this binary inclusion or exclusion of a branch from the eigenstate's support is only an approximation.  As we will discuss below, non-ergodic eigenstates actually decay exponentially along all of the branches.  The rate of decay varies from branch to branch.  We are interested in the number of branches which decay ``sufficiently slowly'', as a function of distance.  There are $N_d = 3 \cdot 2^d \sim 2^d$ sites at a distance $d$ from any given node, so, roughly, if $|\psi_d|^2$ decays more slowly than $2^{-d}$, the normalization sum
\begin{equation}
    \braket{\psi} \approx \sum_{d = 0}^\infty N_d |\psi_d|^2
\end{equation}
diverges.  In the non-ergodic extended state, the wavefunction can be thought of as extending along, instead of all $N_d \sim 2^d$ branches, only $C_d \sim n^d$ of them, where $1 < n < 2$.  For ergodic states, $n = 2$.  Finally, for sufficiently high disorder, $W > W_c$, the state is localized, normalizable, and does not extend along any of branches, \emph {i.e.}~$n < 1$.  This branching ratio $n$ is connected to the fractal dimension $D$ of the eigenstate by the relation $n = 2^D$.~\cite{Kravtsov18}

In terms of the level repulsion metric from the Wegner flow, $\Xi$~\cite{Savitz17}, we find that ergodic, random matrix behavior occurs when
\begin{equation}
    \Xi_d \gtrsim \Xi_d^\star \sim N_d^{-2} \sim 4^{-d}, \label {XiStar}
\end{equation}
where $\Xi_d$ refers to only those $\Xi_{ab}$ values where $a$ is at a distance $d$ from $b$, and $\Xi_d^\star$ is a critical repulsion level at that distance.  This is the square of the amount of energetic level repulsion necessary to shift initially Poissonian eigenvalues with a level density proportional to $N_d$ to a Wigner--Dyson distribution.  This ansatz is confirmed by the numerical results at low disorder in the next section, where our finite system approaches asymptotic ergodicity and reproduces this scaling behavior.

\begin{figure} [bth]
    \begin {center} 
        \includegraphics{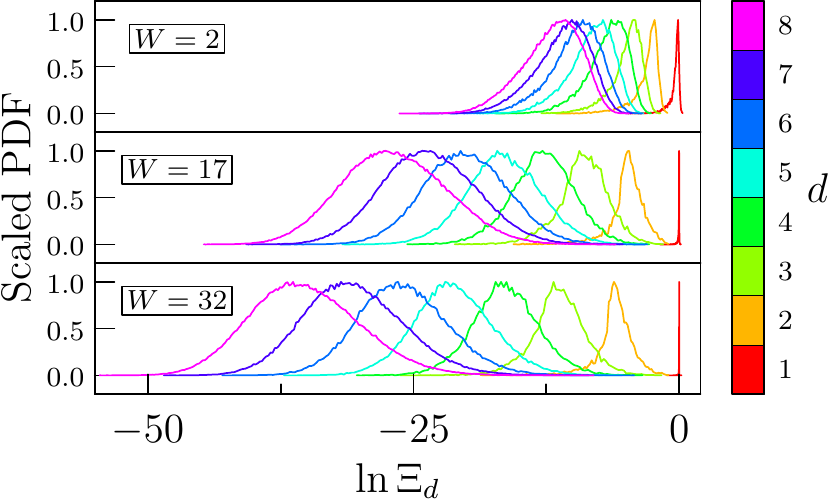}
    \end{center}
    \vspace*{-10pt}
    \caption{
        \textbf{Log-Normal Decay and Spreading of the Level Repulsion Metric with Distance:}
        Sites at further distances interact less in a disorder-dependent manner.  As one moves from right to left, the distribution of the level repulsions of nearest neighbors (red), next-nearest neighbors (orange), etc., and those at distance $d = 8$ (violet), are shown.  For clarity, the bell-shaped curves have been scaled such that their maxima are all fixed at unity.  However, the number of site-pairs at a given distance increases as $N_d \sim 2^d$.  Sites have weaker and more variable interactions at longer distances, and this effect is enhanced by disorder.  Sites which are too close have non-log-normal distributions, and those which are too far apart are influenced by the cage's many long loops, so the equally weighted linear fits in figure~\ref{LinearFits} use only the geometric means and variances of those site-pairs at distances $3 \le d \le 7$.
    } \hrule \vspace*{-5pt} \label{XiPDFs}
\end{figure}

\begin{figure} [bth]
    \begin {center} 
        \includegraphics{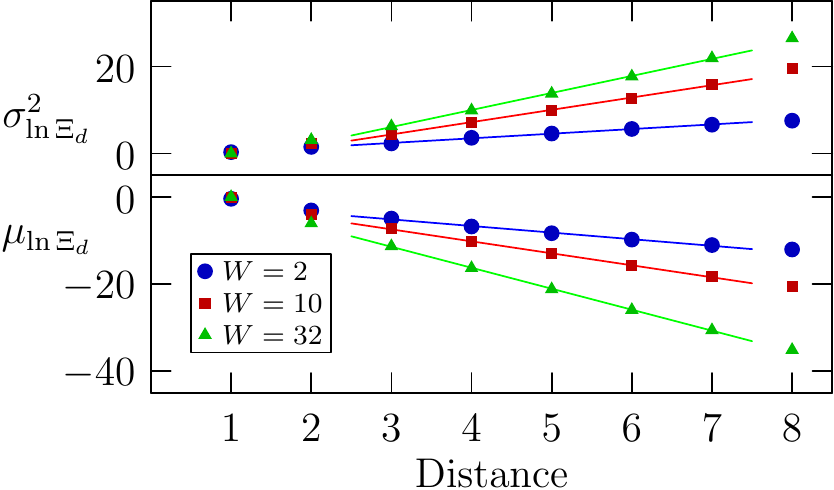}
    \end{center}
    \vspace*{-10pt}
    \caption{
        \textbf {Example Level Repulsion Linear Fits:}
        For intermediate distances, the bulk of the level repulsion data in figure~\ref{XiPDFs} obeys a log-normal distribution with geometric means and variances which grow linearly with distance.  For each disorder strength $W$, we extract $M$ and $S^2$, the slopes of these lines over distance, using an equally weighted linear fit on the data points at distances $3 \le d \le 7$.  In figures~\ref{Results} and~\ref{FractalDimension}, these values are used to classify the behavior of the systems into the three phases shown in figure~\ref{Trees}.
    } \hrule \vspace*{-5pt} \label{LinearFits}
\end{figure}

In other words, equation~\eqref {XiStar} quantifies the reasonable claim that sites at further distances interact less, even for the clean $W = 0$ tree, and the level repulsion behavior of systems in this ergodic class is given by that equation.  For non-ergodic states, the decay of $\Xi$ as one moves away from a central node can be thought of in probabilistic terms.  Each step away from the center approximately multiplies the level repulsion between the center and its parent node by some random variable.  On small scales, the two child nodes of a single parent do interact while diagonalizing the Hamiltonian, but on longer scales ($d \gtrapprox 3$), the geometric central limit theorem kicks in and it becomes reasonable to approximate the $\Xi_d$ distribution as log-normal.  This intuition is borne out in figure~\ref{XiPDFs}.
Furthermore, as $d$ increases, the geometric means and variances of the $\Xi_d$ grow linearly in accordance with the geometric central limit theorem:  $\mu_{\ln {\Xi_d}} \sim M d$ and $\sigma_{\ln {\Xi_d}}^2 \sim S^2 d$.  This linear behavior is demonstrated in figure~\ref{LinearFits}.

\subsection{Caveats}
For small $d \lessapprox 2$, the geometric central limit theorem is clearly not applicable, and their distributions in figure~\ref{XiPDFs} are not bell-shaped.  Furthermore, for large $d$ comparable to half the cage's girth $g/2 = 9$, sites are connected by multiple paths of similar lengths, leading to a breakdown in the cage's ability to approximate the Bethe lattice.  We therefore used only distances $3 \le d \le 7$ for the linear fits shown in figure~\ref{LinearFits}.  Due to our purely numerical approach, we can not assess the complete ergodicity or the possibility of very large crossover scales with certainty, although an effort was made to extrapolate the results from our finite cage to the infinite Bethe lattice below.

This extrapolation will involve examining exponentially small tails of the log-normal distributions given by the geometric central limit theorem.  Importantly, the central limit theorem only applies to these exponentially small quantiles up to a constant factor.  This is unimportant for our qualitative picture, but does provide a degree of uncertainty to the precise values of the critical disorder values $W_E$ and $W_c$ that we obtain. For example, the upper ${2^{-N}}^\textrm{th}$ quantile of the sum of $N$ independent centered Bernoulli distributions is $N$, but the central limit theorem approximation gives $\sqrt{\ln 4} \, N \approx 1.177\,N$.

Finally, we perform our calculations thrice.  Once without regard to the eigenstate energies and once for only $\Xi_{ab}$ values where $|E_a| \le 1/2$.  Without this band resolution, we consider all of the eigenstates together, and can not expect to see ergodicity for positive disorder, as states in edges of the band are likely non-ergodic (although potentially still extended~\cite{Aizenman11,Warzel12,Aizenman13}).  It is only with a restriction to these states in the center of the band that we can possibly expect to find the three phases and the correct critical disorder values.  The third calculation is done without band resolution, but using only the $\Xi_d$ data from distances $d = 3$, $4$, and $5$, in order to give a sense of how the results scale.  With these warnings out of the way, we think that the following approach still provides a clear intuitive picture of the three predicted regimes of behavior.

\subsection{Mathematical Analysis}
In our log-normal approximation, the probability distribution $P(\ln \Xi_d) \propto e^{-z_d^2 /2}$, where 
\begin {equation}
    z_d = \frac {\ln \Xi_d - M d} {S \sqrt d}.
\end {equation}
The number of ``connected'' sites at a distance $d$, $C_d \approx N_d P \left( \Xi_d \ge \Xi_d^\star \right)$, where $\Xi_d^\star$ was defined in equation~\eqref{XiStar}.  Asymptotically, this is proportional to $2^d \erfc(z_d^\star/\sqrt 2)$, where $\erfc$ is the complementary error function, $z_d^\star = - Q \sqrt {d}$, and
\begin{equation}
    Q = \frac {M + \ln 4} S.
\end{equation}
As mentioned above, as $d \to \infty$, $C_d \sim n^d$.  Expanding the tail of the error function gives
\begin{equation}
    C_d \propto 2^d e^{-{z_d^\star}^2 \! /2} = e^ { \left( \ln 4 - Q^2 \right) \frac d 2},
\end{equation}
so we can conclude that the branching ratio
\begin{equation}
    n \approx 2 e^{- Q^2 \! /2 }, \label {n}
\end{equation}
and the fractal dimension
\begin{equation}
    D = \log_2 n \approx 1 - \frac {Q^2} {\ln 4}. \label{D}
\end{equation}

Ergodic states must be entirely connected, \emph{i.e.}~$C_d \sim 2^d$, and so for them, $n = 2$.  This only occurs when $M = -\ln 4$.  We observe this behavior as $W \to 0^+$ and for positive disorder strengths when we restrict our attention to the center of the band.  Equation~\eqref{n} predicts that non-ergodic extended states with $1 < n < 2$ occur when $M < - \ln 4$ and $Q > -\sqrt{\ln 4}$ or
\begin{equation}
    S^2 > \frac { \left( M + \ln 4 \right) ^2} {\ln 4} = \frac {M^2} {\ln 4} + 2M + \ln 4.  \label{LocalizedBound}
\end{equation}
Finally, in the localized phase, this inequality reverses, giving $n < 1$.  Extreme localization occurs as $W \to \infty$, $Q \to -\infty$ and $n \to 0^+$.  All of these types of behavior are observed in the numerical data in the next section.

\section {Numerical Results} \label{NumericalResults}
\begin{figure} [bth]
    \begin {center} 
        \includegraphics{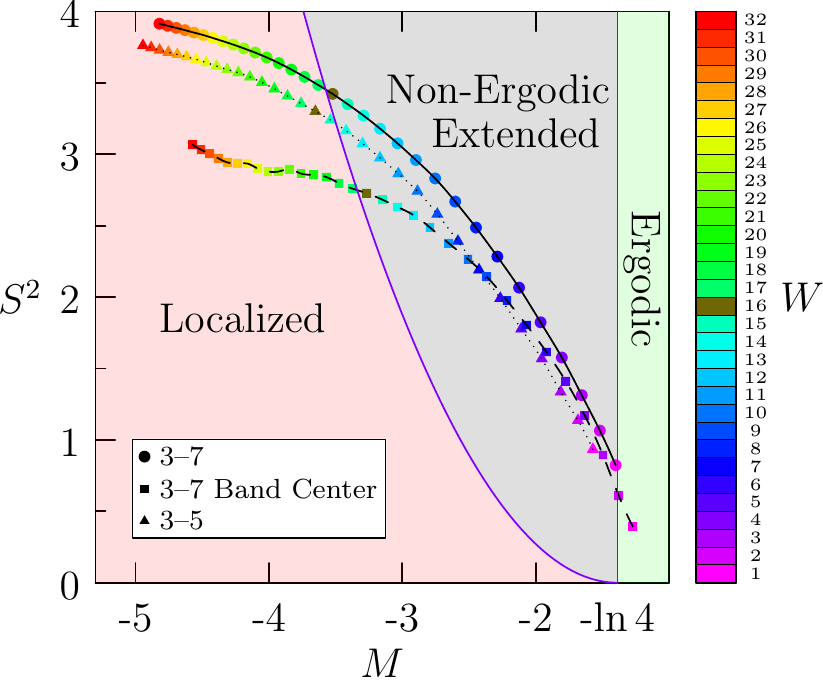}
    \end{center} 
    \vspace*{-10pt}
    \caption{
        \textbf {Level Repulsion Linear Fit Results:}
        The fitted decay and spreading rates of the level repulsion data from figure~\ref{LinearFits} are plotted for integer disorder values $1 \le W \le 32$.  After extrapolating to infinite distance, assuming the validity of the geometric central limit theorem, one can conclude that those systems to the left of the parabola are localized, those within it are extended, and those on its axis with $M = -\ln 4$ are ergodic.  One should not expect to find exact values for $W_c$ with this approach due to the invalidity of the central limit theorem for exponentially small tails.  Additionally, only the band-resolved data could be expected to give precise $W_c$ values or full ergodicity for weak disorder.
    } \hrule \vspace*{-5pt} \label{Results}
\end{figure}

Examples of the $\Xi$ data and its log-normal parameter fits were presented in figures~\ref{XiPDFs} and~\ref{LinearFits} above.  In figure~\ref {Results}, the values obtained for $M$ and $S^2$ are plotted as $W$ varies.
For the data without band resolution, the ergodic $M = -\ln 4$ vertical line is only achieved as $W$ approaches zero and the system becomes clean.  The non-ergodic extended regime, lying between the parabola from equation~\eqref{LocalizedBound} and the ergodic line, occurs for $W \lessapprox 16.5 \approx W_c$.  At the band center, $M \approx -\ln 4$ for $W \lessapprox 2.1 \approx W_E$, supporting the existence of ergodic behavior with positive disorder, and non-ergodic extended states are found for $W_E \approx 2.1 \lessapprox W \lessapprox 16.7 \approx W_c$.  Using only data for $3 \le d \le 5$ (without band resolution) gives $W_c \approx 14.6$.  Localized behavior occurs to the left of the parabola for higher disorder values.

As mentioned in the previous sections, we do not expect this method to be optimally efficient for estimating these critical disorder strengths, and the failure of the central limit theorem in exponentially small tails means that they should not even converge to the true values when applied to increasingly large cages.  The qualitative conclusions remain reasonable, and the estimates are quantitatively in the same ballpark as have been proposed by others.
    
\begin{figure} [bth]
    \begin {center} 
        \includegraphics{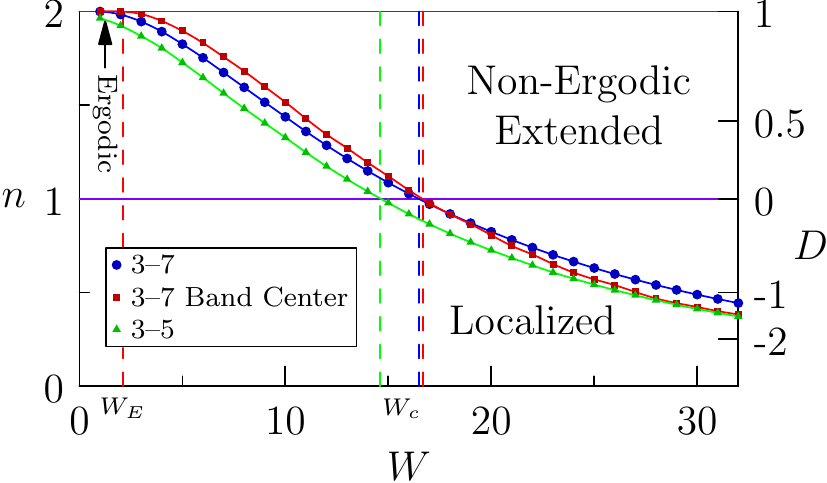}
    \end{center} 
    \vspace*{-10pt}
    \caption{
        \textbf {Effective Branching Rates:}
        The $M$ and $S^2$ data in figure~\ref{Results} can be recast in terms of an effective branching rate $n$ according to equation~\eqref{n} or the fractal dimension $D$ according to equation~\eqref{D}.  At a distance $d$ from a given site, there are $N_d \sim 2^d$ neighbors.  According to our binary approximation, we say that the site is ``connected'' to $C_d \sim n^d$ of them in the infinite distance limit.  Ergodic extended behavior is recovered when $n = 2$ and essentially all site pairs are connected.  For $1 < n < 2$, the number of connected nodes grows exponentially with distance, but they constitute a vanishing fraction of the tree, giving the non-ergodic extended phase.  Finally, for $n < 1$, no site-pairs are connected at large distances, and the behavior is localized.  Note that for weak disorder, particularly at the center of the band, numerical results gave some $M$ slightly above $-\ln 4$.  We assume these data points correspond to the ergodic phase and result from the finite size of the cage, and so we set them to $n = 2$.
    } \hrule \vspace*{-5pt} \label{FractalDimension}
\end{figure}

One can evaluate the the branching ratios $n$ and fractal dimensions $D$ as a function of $W$ by applying equations~\eqref{n} and~\eqref{D} to the $M$ and $S^2$ found for each disorder strength.  Thus, in figure~\ref{FractalDimension}, the top line at $n = 2$ corresponds to ergodicity, and that at $n = 1$ is the transition from the non-ergodic extended phase to localization.

\section {Discussion} \label{Discussion}
Despite the apparent simplicity of the disordered Bethe lattice, its intricacies have continued to astonish the mathematicians and physicists who study it for decades.  Reviewing the literature, there are repeated instances of researchers who seem to have ``closed the book'' on this model, only for new details to emerge which either indicate that its behavior is richer than anticipated or even contradict the existing understanding.  In the words of Simone Warzel, there are many ``surprises in the phase diagram of the Anderson model on the Bethe lattice~\cite{Warzel12}'', and we can only guess when the surprises will come to a final conclusion.

This confusion is due to a number of factors.  This model's behavior is indeed more complex than one would na{\"\i}vely guess based on finite-dimensional localization results.  Finite-dimensional Anderson localization is already difficult to experimentally observe, but direct physical embeddings of the infinite-dimensional Bethe lattice are even less practical to construct, and so there is little hope of experimental data resolving the situation.  Achieving consensus has also been made more difficult due to the variation and inconsistencies in terminology used between mathematicians studying spectral theory, theoretical physicists, and computational physicists.  As reviewed in section~\ref{Review}, a wide range of advanced approaches have been employed to tackle this model by all three of these groups.  The numerical situation is complicated by the necessity of selecting a finite version of the Bethe lattice, and it is not always clear how the various choices affect the model's behavior in the scaling limit.~\cite {Ostilli12}  It is difficult to rigorously define extended states on infinite graphs, and it is not clear which if any of the finite graphs converge to the infinite Bethe lattice limit as they grow in size.  Numerical approaches are also hindered by the exponential growth of the number of sites of the Bethe lattice with its linear dimension, thus limiting our ability to test the existence of a large non-ergodic extended--ergodic phase crossover length scale such as that claimed in~\cite{Tikhonov16a,Tikhonov16b,Biroli17,Biroli18,Metz17,Tikhonov19,Tikhonov19b}.  It seems possible that random regular graphs and cages exhibit such a crossover due to their exponentially many large loops, but the truly loopless infinite Bethe lattice does not.  In that case, we hope that our extrapolation to infinitely distant sites in section~\ref {Picture} gives results which are more applicable to the infinite Bethe lattice.

Another possible explanation for the discrepancy between our results and the growing belief in the nonergodic extended--ergodic phase crossover goes as follows:  At a significant distance from an arbitrarily chosen central node, amongst the exponentially growing number of sites, perhaps one site has a particularly small energetic level spacing from the eigenstate developing at our central node when compared to the coupling generated between them by the flow.  While the flow would then perform a large rotation in Hilbert space, possibly leading to ergodicity, the level repulsion metric $\Xi$ would not reflect this, as the weak coupling would cause the levels to repel only very slightly.  The reasoning presented in this work is based on the commonly held belief that ergodic states generally exhibit Wigner--Dyson level statistics.  We therefore tentatively suggest that, if these ergodic crossover states truly exist on the infinite tree, they may provide a counterexample to this dictum.  By our understanding, the analysis using the $\Xi$ metric above indicates that these ultimately ergodic crossover states should have Poissonian level statistics.

Despite the difficulties mentioned above, the importance of understanding the disordered tree is growing as comparisons between it and the explosive field of many-body localization continue to be drawn.  Tests of this correspondence would be quite valuable.  We considered trying to apply the techniques of this paper to a real-space MBL system, but were unable to find a numerically tractable model with a satisfactory distance metric.  Success in this endeavour would be a notable contribution to the MBL literature, as the possibility of non-ergodic states extended in Fock space has been often overlooked.

Some additional directions that might extend the approach in this paper include larger cages, a more traditional measure of eigenstate overlap instead of the $\Xi$ metric from the flow, a more careful mathematical analysis of the exponentially small tails of the log-normal distributions (such as in \cite{Hoecker14}), a more elaborate approach for isolating the band center, a comparison between the scaling limits of cages and random regular graphs, and an assessment of whether the arguments for a large non-ergodic extended--ergodic phase crossover length scale on random regular graphs~\cite{Tikhonov16a,Tikhonov16b,Biroli17,Biroli18,Metz17,Tikhonov19,Tikhonov19b} extends to our approach.  One might expect more precise values for the critical disorder strengths $W_E$ and $W_c$ to emerge if these improvements were made.

We hope that our introduction of a new approach and setting, namely Wegner flows and cages, does not merely accentuate the aforementioned confusion.  Further graph settings that may be of interest are those with quenched topological disorder in spaces of constant negative curvature such as ``hyperbolic geometric graphs''~\cite{Krioukov10} and Delaunay triangulations of randomly placed points on the hyperbolic plane~\cite{Bogdanov14}.  Due to their large boundary, it may be difficult to construct a finite analog of such graphs for numerical purposes, but means for resolving such issues in percolation theory have been proposed~\cite{Sausset10}.  The topological disorder may obviate the need for on-site disorder, possibly giving \emph{two critical delocalizing negative curvatures} for a given \emph{dimensionless} model of topological disorder at $W = 0$ and unit site density, but numerical evidence on $\mathbb R^2$~\cite{Puschmann15} indicates that the localization lengths are likely to be prohibitively large.  The $\mathbb H^2$ case, or the AdS$_{2 + 1}$ setting for dynamical studies, seems most promising, as any amount of topological disorder should localize all states in the weak curvature limit.

The Anderson model on the Bethe lattice may be connected to more than just MBL.  The Bethe lattice can be viewed as a discretization of hyperbolic space which has a $p$-adic boundary at infinity.  Perhaps there is an equivalent formulation in terms of a theory with power-law correlations on this $p = 2$-adic boundary, akin to the AdS/CFT holographic paradigm in high-energy physics.~\cite{Maldacena99,Gubser17,Heydeman16}  The power law exponent depends on the strength of the localization in the hyperbolic bulk.  Interestingly, power law correlator decay shows two distinct regimes of behavior depending on whether the implied theories exhibit IR and/or UV divergences, which may correspond to the non-ergodic extended and localized phases in the bulk, respectively.  More speculatively, one might hope to connect this with Kehrein's ``flow equation holography'' proposal.~\cite{Kehrein17}

\begin {acknowledgments}
    Thanks to Evert van Nieuwenburg, Yuval Baum, Stefan Kehrein, and Matthew Heydeman for fruitful discussions, and to Konstantin Tikhonov, Gabriel Lemari\'e, Steven Thompson, and our two anonymous referees for their feedback and suggestions regarding our preprint.
    
    This work was supported by the Institute for Quantum Information and Matter (IQIM), a National Science Foundation (NSF) frontier center partially funded by the Gordon and Betty Moore Foundation.  S.S.~was funded by Grant No.~DGE-1745301 from the NSF Graduate Research Fellowship.  C.P.~thanks the Caltech Student--Faculty Programs office and the Blinkenberg family for their support.  G.R.~acknowledges the generous support of the Packard Foundation and the IQIM.
    
    The numerical Wegner flows were implemented using floating-point matrices calculated by the open-source linear algebra library \textsc{Armadillo}.~\cite{SandersonCurtin16}
\end {acknowledgments}

\bibliography {library}

\end {document}